\documentclass{emulateapj}
\usepackage{apjfonts}
\usepackage{psfig,epsfig}

\def\spose#1{\hbox to 0pt{#1\hss}}
\def\lta{\mathrel{\spose{\lower 3pt\hbox{$\sim$}}
    \raise 2.0pt\hbox{$<$}}}
\def\gta{\mathrel{\spose{\lower 3pt\hbox{$\sim$}}
    \raise 2.0pt\hbox{$>$}}}
\begin{document}

\title{A correlation between light profile and [Mg/Fe] abundance
ratio in early--type galaxies}
\author{Alexandre Vazdekis \altaffilmark{1}, Ignacio Trujillo\altaffilmark{2}
and Yoshihiko Yamada \altaffilmark{3}}
\affil{ $^1$ Instituto de Astrof\'{\i}sica de Canarias, E-38200 La
Laguna, Tenerife, Spain; vazdekis@ll.iac.es} \affil{ $^2$
Max--Planck--Institut f\"ur Astronomie, K\"onigstuhl 17, D--69117
Heidelberg, Germany }
\affil{ $^3$
National Astronomical Observatory of Japan, 2-21-1 Osawa, Mitaka, Tokyo 181-8588, Japan}


\begin{abstract}

We explore possible correlations between light profile shapes, as parameterized
by the S\'ersic index $n$ or the concentration index $C_{re}(1/3)$, and
relevant stellar population parameters in early-type galaxies.  Mean luminosity
weighted ages, metallicities and abundance ratios were obtained from spectra of
very high signal--to--noise and stellar population models that synthesize
galaxy spectra at the resolution given by their velocity dispersions $\sigma$,
in combination with an age indicator (H$\gamma_{\sigma}$) that is virtually
free of the effects of metallicity. We do not find any significant correlation
between $n$ (or $C_{re}(1/3)$) and mean age or metallicity, but a strong
positive correlation of the shape parameters with Mg/Fe abundance ratio. This
dependence is as strong as the Mg/Fe -- $\sigma$ and $C_{re}(1/3)$ -- $\sigma$
relations. We speculate that early-type galaxies settle up their structure on
time-scales in agreement with those imposed by their Mg/Fe ratios. This suggest
that the global structure of larger galaxies, with larger Mg/Fe ratios and
shorter time-scales, was already at place at high $z$, without experiencing a
significant time evolution. 

\end{abstract}

\section{Introduction}
\label{Sec:intro}

A successful galaxy formation model should be able to explain the main scaling
relations followed by early-type (E/SO) galaxies. These galaxies show up a
tight relation between the size (by means of the effective radius $r_e$),
surface brightness, and central velocity dispersion, $\sigma$, which is known
as the Fundamental Plane (Djorgovski \& Davis 1987; Dressler et al. 1987).
Early-type galaxies also follow the Color-Magnitude Relation, CMR, i.e.
brighter galaxies look redder (Visvanathan \& Sandage 1977; Bower et al. 1992).
The CMR and the Mg$_2$--$\sigma$ relation (e.g. Bender, Burstein \& Faber 1992;
Colless et al. 1999) links the mass of a galaxy, through its luminosity, to its
constituent stellar populations. However, little is known about the relation
between the morphological shape and the stellar populations of early-type
galaxies (Conselice 2003), most likely because these two fields rely on methods
that have been traditionally disconnected.

From a stellar population point of view, measured colors and absorption
line-strengths are translated to mean ages and metallicities via stellar
population synthesis models. Using these models the CMR is often interpreted as
a metallicity sequence (e.g., Arimoto \& Yoshii 1987), i.e. elliptical galaxies
are old and passively evolving systems (e.g., Ellis et al. 1997; Stanford,
Eisenhardt \& Dickinson 1998). However, this view is questioned by some
studies, which show evidence for a significant intermediate age population in
some ellipticals (e.g., Gonz\'alez 1993; J\o rgensen 1999).  Element ratios
provide further constraints on the star formation histories (e.g. Worthey
1998). These studies have shown compelling evidence for a systematic departure
from solar element ratios of some elements as a function of $\sigma$ (Trager et
al. 2000). Recent modeling efforts have predicted spectra of stellar
populations at moderately high resolution (Vazdekis 1999; Schiavon et al. 2002;
Bruzual \& Charlot 2003), leading to more accurate stellar population parameter
estimates (Vazdekis \& Arimoto 1999; Vazdekis et al. 2001a; Kuntschner et al.
2002).

From the point of view of galaxy structure, the shape of the overall surface
brightness as parametrized by S\'ersic (1968) index, $n$, correlates with the
global properties of early-type galaxies: total luminosity, central surface
brightness and effective radius, $r_e$ (Graham, Trujillo \& Caon 2001 and
references therein). Nowadays it is clear that the above luminosity--dependent
departures from de Vaucouleurs profile are real, as clearly proven by the
existence of strong correlations between $n$ (or equivalently galaxy central
concentration of light $C_{re}(1/3)$ (Trujillo, Graham \& Caon 2001)) and the
photometric--independent properties,  $\sigma$ (Graham, Trujillo \& Caon 2001)
and galaxy's supermassive black hole mass (Graham et al. 2001b).

Aiming at exploring possible correlations between galaxy light profiles and
their stellar populations we combine in this letter the results obtained from
applying these modeling developments to high quality spectroscopic
data. We find a tight correlation between [Mg/Fe] ratio and the
shape parameters.

\begin{deluxetable*}{rrrrrrrrrr}
  \tablecaption{Galaxy measurements \label{tbl-1}} \tablewidth{0pt}
  \tablehead{\colhead{NGC}&\colhead{n}&\colhead{Ref}&\colhead{$C_{re}(1/3)$}&\colhead{$\sigma$(${\rm km~s^{-1}}$)}&
\colhead{Age (Gyr)}& \colhead{[M/H]}& \colhead{[Fe/H]}& \colhead{[Mg/H]}& \colhead{[Mg/Fe]} }
  \startdata
  \multicolumn{10}{c}{Virgo galaxies}\\[4pt]
4239 & 0.67 & 2 & 0.17 &  82 &  5.5$^{+ 1.9}_{-1.7}$&-0.21$^{+0.11}_{-0.10}$& -0.21$^{+0.09}_{-0.09}$& -0.16$^{+0.15}_{-0.14}$& 0.05$^{+0.10}_{-0.09}$ \\[2pt]
4339 & 3.50 & 1 & 0.44 & 142 & 12.6$^{+ 5.3}_{-2.8}$&-0.06$^{+0.10}_{-0.06}$& -0.17$^{+0.07}_{-0.07}$& +0.13$^{+0.15}_{-0.15}$& 0.30$^{+0.11}_{-0.11}$ \\[2pt]
4365 & 6.08 & 1 & 0.55 & 245 & 20.4$^{+ 3.1}_{-6.4}$&-0.02$^{+0.10}_{-0.08}$& -0.28$^{+0.16}_{-0.08}$& +0.21$^{+0.21}_{-0.16}$& 0.49$^{+0.08}_{-0.11}$ \\[2pt]
4387 & 1.92 & 1 & 0.34 & 105 & 15.1$^{+ 5.4}_{-4.8}$&-0.20$^{+0.11}_{-0.11}$& -0.22$^{+0.10}_{-0.08}$& -0.09$^{+0.19}_{-0.10}$& 0.13$^{+0.12}_{-0.05}$ \\[2pt]
4458 & 2.55 & 1 & 0.40 & 105 & 16.6$^{+ 3.9}_{-5.6}$&-0.31$^{+0.13}_{-0.05}$& -0.40$^{+0.05}_{-0.10}$& -0.05$^{+0.16}_{-0.07}$& 0.35$^{+0.14}_{-0.03}$ \\[2pt]
4464 & 2.61 & 1 & 0.38 & 135 & 21.0$^{+ 2.3}_{-7.1}$&-0.32$^{+0.12}_{-0.06}$& -0.40$^{+0.09}_{-0.07}$& -0.03$^{+0.12}_{-0.13}$& 0.37$^{+0.06}_{-0.09}$ \\[2pt]
4467 & 1.54 & 2 & 0.28 &  75 & 16.0$^{+ 5.2}_{-5.3}$&-0.23$^{+0.13}_{-0.12}$& -0.35$^{+0.13}_{-0.08}$& -0.08$^{+0.18}_{-0.13}$& 0.27$^{+0.08}_{-0.08}$ \\[2pt]
4472 & 6.27 & 1 & 0.55 & 306 & 10.4$^{+ 9.5}_{-4.1}$& 0.07$^{+0.17}_{-0.20}$& -0.12$^{+0.15}_{-0.23}$& +0.46$^{+0.16}_{-0.37}$& 0.58$^{+0.04}_{-0.17}$ \\[2pt]
4473 & 5.29 & 1 & 0.52 & 180 & 10.3$^{+ 3.1}_{-2.1}$& 0.15$^{+0.07}_{-0.11}$& -0.06$^{+0.06}_{-0.08}$& +0.48$^{+0.08}_{-0.16}$& 0.54$^{+0.04}_{-0.10}$ \\[2pt]
4478 & 1.98 & 1 & 0.33 & 135 & 10.0$^{+ 2.7}_{-2.8}$& 0.04$^{+0.15}_{-0.12}$& -0.08$^{+0.12}_{-0.06}$& +0.23$^{+0.14}_{-0.17}$& 0.31$^{+0.05}_{-0.14}$ \\[2pt]
4489 & 2.50 & 4 & 0.37 &  73 &  3.1$^{+ 0.6}_{-0.8}$& 0.20$^{+0.18}_{-0.16}$& +0.20$^{+0.15}_{-0.15}$& +0.23$^{+0.17}_{-0.21}$& 0.03$^{+0.06}_{-0.10}$ \\[2pt]
4551 & 1.89 & 1 & 0.32 & 105 &  7.2$^{+ 1.5}_{-0.9}$& 0.19$^{+0.07}_{-0.09}$& +0.07$^{+0.05}_{-0.10}$& +0.41$^{+0.07}_{-0.12}$& 0.34$^{+0.05}_{-0.05}$ \\[2pt]
4621 & 6.14 & 1 & 0.55 & 230 & 10.3$^{+ 2.4}_{-1.8}$& 0.26$^{+0.08}_{-0.08}$& -0.04$^{+0.04}_{-0.04}$& +0.64$^{+0.05}_{-0.14}$& 0.60$^{+0.02}_{-0.11}$ \\[2pt]
4697 & 3.70 & 3 & 0.45 & 168 & 11.3$^{+ 4.5}_{-2.7}$& 0.06$^{+0.10}_{-0.11}$& -0.10$^{+0.06}_{-0.10}$& +0.30$^{+0.08}_{-0.17}$& 0.40$^{+0.04}_{-0.09}$ \\[2pt]
  \multicolumn{10}{c}{Field galaxies}\\[4pt]
 584 & 4.70 & 2 & 0.50 & 217 &  8.6$^{+ 1.9}_{-2.3}$&  -                    & -0.03$^{+0.04}_{-0.08}$& +0.31$^{+0.06}_{-0.10}$&+0.34$^{+0.07}_{-0.05}$ \\[2pt]
 720 & 4.96 & 3 & 0.51 & 247 & 10.9$^{+ 7.0}_{-2.7}$&  -                    & -0.16$^{+0.08}_{-0.08}$& +0.58$^{+0.12}_{-0.19}$&+0.74$^{+0.07}_{-0.14}$ \\[2pt]
 821 & 4.00 & 2 & 0.47 & 199 & 12.3$^{+12.5}_{-5.6}$&  -                    & -0.15$^{+0.17}_{-0.17}$& +0.26$^{+0.18}_{-0.23}$&+0.41$^{+0.04}_{-0.08}$ \\[2pt]
1700 & 5.99 & 2 & 0.54 & 233 &  6.4$^{+ 3.1}_{-1.7}$&  -                    & +0.04$^{+0.08}_{-0.09}$& +0.37$^{+0.09}_{-0.15}$&+0.34$^{+0.09}_{-0.09}$ \\[2pt]
3379 & 4.64 & 3 & 0.49 & 201 & 17.8$^{+ 7.4}_{-6.9}$&  -                    & -0.29$^{+0.08}_{-0.10}$& +0.19$^{+0.19}_{-0.10}$&+0.48$^{+0.12}_{-0.03}$ \\[2pt]
5831 & 4.72 & 3 & 0.50 & 166 &  5.9$^{+ 3.5}_{-1.9}$& 0.32$^{+0.15}_{-0.13}$& +0.14$^{+0.15}_{-0.16}$& +0.56$^{+0.06}_{-0.18}$&+0.42$^{+0.04}_{-0.10}$ \\[2pt]
7454 & 2.55 & 4 & 0.38 & 112 &  6.5$^{+ 3.3}_{-1.8}$&  -                    & -0.20$^{+0.09}_{-0.10}$& -0.07$^{+0.13}_{-0.14}$&+0.13$^{+0.06}_{-0.06}$
  \enddata

\tablecomments{The numbers in column 3 indicate the $n$ parameter sources: 1
Caon, Capaccioli \& D'Onofrio (1993). 2 HST F555W archive images. 3  Fits to
the profiles of Peletier et al. (1990). 4 V--band images from J. Blakeslee
(private communication).} 

\end{deluxetable*}

\section{Galaxy measurements}
\label{Sec:data}

We use the results derived by Vazdekis et al (2001a) and Yamada et al. (2003a,
in prep.) for the stellar populations of 14 early-type galaxies in the Virgo
cluster. We also included here 7 early-type galaxies in lower density
environments (Yamada et al. 2003b, in prep.). Long-slit spectra of extremely
high quality were obtained at the William Herschel (4.2m) and the Subaru (8.2m)
telescopes, covering the range $\lambda\lambda \sim$ 4000-5500~\AA, at
resolution 2.0--3.1\AA~(FWHM). Signal-to-noise ranges from 100 to 500 (S/N per
\AA\ unit), with larger values for galaxies with larger $\sigma$'s, to be able
to accurately measure the new H$\gamma_\sigma$ age indicator of Vazdekis \&
Arimoto (1999). Spectra were flux calibrated and summed up within $r_{e}$/10.
Line-strengths and $\sigma$ were measured from the summed spectra. Note that
half of the sample has $\sigma < {\rm 150~kms^{-1}}$, allowing us to cover a
large galaxy mass range. 

Mean luminosity weighted ages, metallicities and abundance ratios were
estimated by comparing selected absorption line-strengths with those predicted
by the model of Vazdekis (1999). This model provides flux calibrated spectra in
the optical range at resolution 1.8~\AA\/ (FWHM) for single burst stellar
populations (SSPs). An important advantage over previous approaches, which only
predicted the strengths of a number of absorption lines (e.g. Worthey 1994;
Vazdekis et al. 1996), is that we are able to analyze galaxy spectra at the
resolution given by their internal velocity broadening and instrumental
resolution. Therefore we do not work at the resolutions (FWHM$>$8\AA) of the
Lick/IDS system (Worthey 1994), and we do not transform our line-strengths
(direcly measured on the flux-calibrated observational and synthetic spectra)
to the response curve of the IDS spectrograph. 

Plots of the strengths of several metal line indices such as Mg$b$ (Worthey
1994) or Fe3 (Kuntschner 2000) versus H$\gamma_\sigma$, which is virtually free
from metallicity dependence, provide almost orthogonal model grids, allowing us
to accurately estimate galaxy mean ages as well as the abundances of these
elements. For a given galaxy all these plots provide almost identical ages, but
the obtained metallicities will be different if the galaxy shows a departure
from scaled-solar element ratios as in the models. Since the Mg$b$ index is by
far dominated by Mg and the Fe3 index by Fe (Tripicco \& Bell 1995), the
obtained metallicities can be approximated to [Mg/H] and [Fe/H] abundances,
respectively. These values are then used to estimate the [Mg/Fe] abundance
ratio. An extrapolation of the model grids is usually required to obtain the
[Mg/H] for most massive galaxies, which show the largest [Mg/Fe] ratios, since
the models only extend to [M/H]=+0.2. Finally, plotting the [MgFe] index
(Gonz\'alez 1993), which is nearly insensitive to non solar abundance ratios
(Vazdekis et al. 2001a; Thomas, Maraston \& Bender 2003), versus
H$\gamma_\sigma$ allows us to derive the total metallicity [M/H]. An
alternative approach for obtaining the [Mg/Fe] ratios can be followed using
models specifically computed for different $\alpha$-enhancements (e.g. Trager
et al. 2000; Thomas et al. 2003), on the basis of the sensitivities of these
lines to the abundance changes of the different species as tabulated in
Tripicco \& Bell (1995).
It is worth noting that fitting galaxy spectra with single burst models
provides mean luminosity-weighted ages and metallicities, which means that the
presence of young populations dominate the mean age and therefore are easily
detected. Table~{\ref{tbl-1}} summarizes our results.

The values of $n$ for most of Virgo galaxies were taken from Caon, Capaccioli
\& D'Onofrio (1993) based on B--band profiles. For the remaining galaxies we
proceeded as follows: {\it i)} S\'ersic model fitting of published B--band
profiles (Peletier et al. 1990) or {\it ii)} when B--band images were not
available we fitted V or F555W HST light profiles derived from archive images
(see footnote of Table~{\ref{tbl-1}}). Due to the small B-V color gradient of
early-type galaxies (e.g. Peletier et al. 1990), $n$ does not depend much on
the filter in use. For NGC~4239 and NGC~4489 a S\'ersic law was
unable to fit observed light profile and a bulge ($r^{1/n}$) plus disk
(exponential) decomposition was applied and the derived $n$ value corresponds
to the bulge. An alternative way of describing the light profile shape is using
the central concentration index of Trujillo et al. (2001), $C_{re}(1/3)$. This
index, which has been shown to be robust against measurement errors (Graham et
al. 2001a) and monotonically related to $n$, evaluates the fraction of light
enclosed within $r_{e}/3$ divided by the light within $r_e$ (i.e. half of the
total light). We use both $n$ and $C_{re}(1/3)$ indices. Errors for $n$ have a
typical uncertainty of 25\% (Caon, Capaccioli \& D'Onofrio 1993), which
translates to 10--15\% uncertainty in $C_{re}(1/3)$.

\begin{figure}[t]
\centerline{\psfig{file=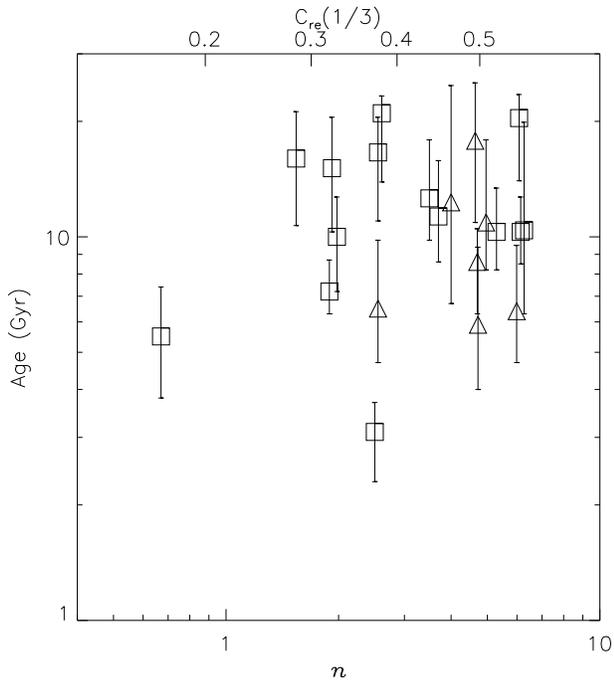,width=3.6in}} 
\caption{
The shape parameters $n$ and $C_{re}(1/3)$
versus mean luminosity weighed ages. Virgo galaxies are represented by open
squares whereas field galaxies are represented by open triangles.
\label{Fig:c1a}}
\end{figure}

\begin{figure}[t]
\centerline{\psfig{file=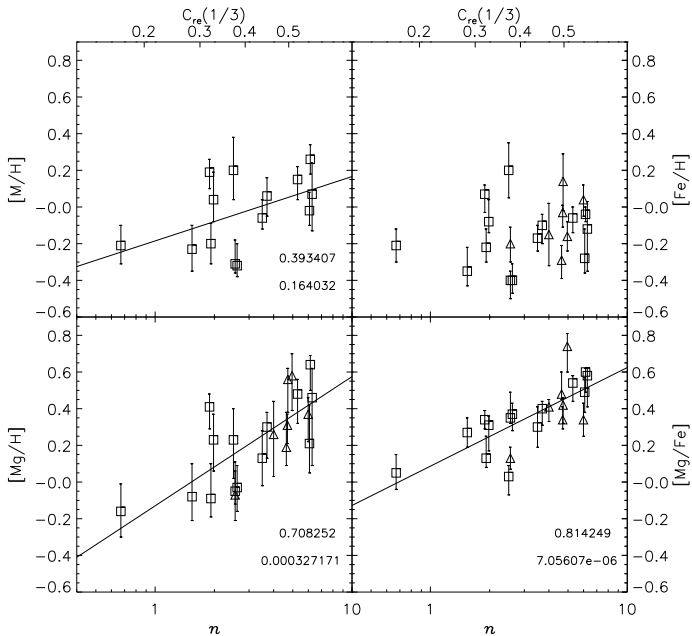,width=3.6in}} 
\caption{
Shape parameters versus mean global metallicity [M/H],  [Fe/H] and
[Mg/H] metallicities and [Mg/Fe] abundance ratio. Symbols have same meaning as
in Fig.~{\ref{Fig:c1a}}. Spearman rank-order correlation coefficient and its
significance value are written in each panel.
\label{Fig:c1m}}
\end{figure}


\section{Results}
\label{Sec:results}

In this section we explore possible correlations between the shape and the
stellar population parameters described in Section~ \ref{Sec:data}. In
Fig.~{\ref{Fig:c1a}} we plot $n$ and $C_{re}(1/3)$ indices versus mean
luminosity weighed ages obtained for Virgo and field early-type galaxies. We do
not see a clear trend of the ages as a function of shape. It is worth noting
that rather than absolute ages, which are subject of model uncertainties, we
are interested in relative age trends (see Vazdekis et al. 2001b and Schiavon
et al. 2002 for a detailed discussion on this issue).

We now focus on metallicities and abundance ratios. First panel of
Fig.~{\ref{Fig:c1m}} shows the total metallicity [M/H] versus $n$ and
$C_{re}(1/3)$. Although we see that more metal-rich galaxies show higher
$C_{re}(1/3)$ values the obtained correlation is weak, as shown by the Spearman
test $r_s$. One would expect that larger galaxies with higher $C_{re}(1/3)$
(according to Graham, Trujillo \& Caon 2001) and stronger potential wells would
supply more fuel to their inner regions increasing their metallicity. This
expectation might not be supported by our results.

Fig.~{\ref{Fig:c1m}} plots $n$ and $C_{re}(1/3)$ versus Mg and Fe
metallicities, showing no trend with Fe but a positive correlation with Mg. The
latter can be understood if we combine the $C_{re}(1/3)$ -- $\sigma$ (Graham,
Trujillo \& Caon 2001) and Mg$_2$ -- $\sigma$ (Bender et al. 1992; Colless et
al. 1999) relations. No full agreement has yet been reached for the Fe --
$\sigma$ relation: J\o rgensen (1999) finds that Fe abundance does not vary
with $\sigma$, whilst Kuntschner (2000) finds a positive trend. This
discrepancy can be in part attributed to the use or not of models including
$\alpha$-enhancement prescriptions (Thomas et al. 2003).

Fig.~{\ref{Fig:c1m}} shows for the first time a strong positive correlation
between [Mg/Fe] abundance ratio and shape parameters. This is expected since
[Mg/Fe] correlates with $\sigma$ (Trager et al. 2000; Kuntschner 2000; Thomas,
Maraston \& Bender 2002), and $\sigma$ correlates with $C_{re}(1/3)$ (Graham,
Trujillo \& Caon 2001). Nonetheless, it is surprising to see that [Mg/Fe]
provides the strongest correlation among other relevant stellar population
parameters, reaching a Spearman coefficient 0.8. A similar correlation strength
is obtained by removing the galaxy with the smallest $n$ value (i.e. NGC~4239).

\section{Discussion}
\label{Sec:discussion}

\begin{figure}[t]
\centerline{\psfig{file=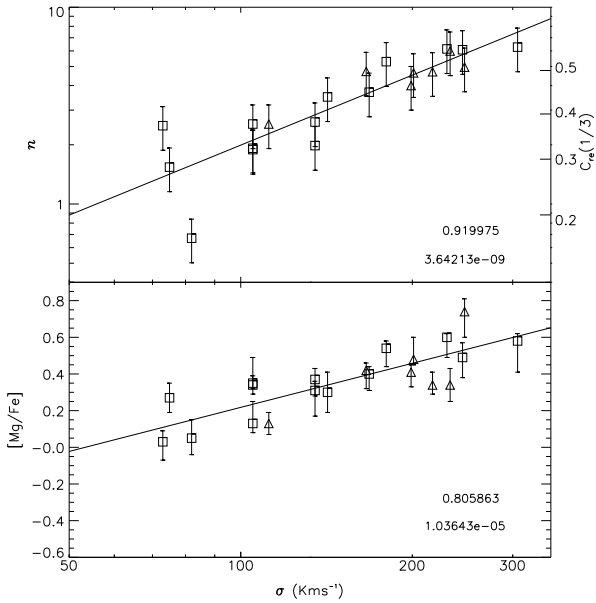,width=3.6in}} 
\caption{Galaxy velocity dispersion versus $n$ and $C_{re}(1/3)$ and [Mg/Fe].
Symbols have same meaning as in Fig.~{\ref{Fig:c1a}}
\label{Fig:c1s}}
\end{figure}

To test how strong is the new correlation between the shape parameters and
[Mg/Fe], we show in Fig.~{\ref{Fig:c1s}} the fits obtained for the already
known $C_{re}(1/3)$ -- $\sigma$ and [Mg/Fe] -- $\sigma$ relations. We use the
same galaxies plotted in Fig.~{\ref{Fig:c1a}} and Fig.~{\ref{Fig:c1m}}. The
[Mg/Fe] -- $n$ relation shows a correlation value as strong as the ones plotted
in Fig.~{\ref{Fig:c1s}}.

It is well known that early-type galaxies show non-solar abundance ratios
(e.g., Trager et al. 2000). The [Mg/Fe] ratio, which spans a
super-solar range (see Fig.~{\ref{Fig:c1m}}), provides us with important clues
on the star-formation time-scales. The $\alpha$ elements, such as Mg, are
ejected to the ISM by SN~II, which explode on short time-scales ($\le10^6-10^7$
yr). On the other hand, SN~Ia are the main producers of Fe peak elements, which
require time-scales around $\sim$1~Gyr. It has been proposed that if the
duration of the star formation is shorter than SN~Ia time-scales the [Mg/Fe]
should be larger than 0 (e.g., Worthey et al. 1992). Other
possibilities have also been proposed, such as a top-heavy initial mass
function (IMF), which could also be combined with a star formation that stopped
before the products of SN~Ia could be incorporated into the stars (e.g.,
Worthey et al. 1992; Vazdekis et al. 1996).

The fact that  the central ($r_{e}/10$) $\sigma$ and [Mg/Fe] measurements are
strongly correlated (see lower panel of Fig.~{\ref{Fig:c1s}}) suggests that
the innermost regions of the more massive galaxies, with larger potential
wells, stopped their star formation on shorter time-scales. Furthermore since
the light concentration index strongly correlates with [Mg/Fe] suggests that
early-type galaxies settled up their global (and not only the innermost
regions) structure on time-scales according to their [Mg/Fe] ratios. This
scenario would be reinforced if no [Mg/Fe] radial gradients were present.
Almost nothing is known about [Mg/Fe] gradients outward galaxy effective
radii, there are indications that this ratio is roughly constant at least
out to 1~$r_e$ (Halliday 1999; Mehlert et al. 2003). Mehlert et al. suggest
that the constraint of short formation time-scales is a global feature and
not only applies to galaxy centers. 

We do not see a clear correlation between $n$ and galaxy mean luminosity
weighted ages but [Mg/Fe] suggesting that the formation time-scale, rather than
burst peppering as a function of time, is the main process linked to galaxy
global structure. Since our galaxies are old and the time scales of the larger
galaxies very short, the new correlation suggests that the global structure of
these galaxies was already at place at high $z$, without experiencing a
significant time evolution. This result is supported by Chiosi \& Carraro
(2002) models, which suggest that the gravitational potential well (and
therefore mass and shape of the galaxy) dictates the efficiency and extention
of the star formation. The favoured picture is that early-type galaxies are
assembled at high $z$ (e.g., Fassano et al. 1998; van Dokkum \& Ellis 2003),
defining a tight sequence in the CM diagram at least out to $z \sim$1.2
(Blakeslee et al. 2003), and their stellar populations evolve passively
(e.g., Im et al. 2002; Schade et al. 1999). Furthermore, Mobasher et al. (2003)
find that early-type galaxies show higher rest-frame B-band concentration
indices than late-type spirals out to $z \sim 1$, which suggest that by that
redshift value early-type systems have already developed large central light
concentrations. Such high light concentrations are predicted by both the
monolithic scenario by means of a rapid gas collapse which turns into stars,
and within hierarchical galaxy formation framework via mergers of stellar and
gaseous systems that take place at high z. Conselice (2003) shows that the
light concentration is a key parameter to trace the past evolutionary
history of galaxy formation.

This research will benefit from studies reaching deeper gradients of
line-strengths and abundance ratios, and from theoretical efforts linking them
to the light profiles (Angeletti \& Giannone 2003). Including other element
ratios in this analysis would provide further clues, since each element is
expelled to the ISM at different epochs, and because line-strengths do not only
depend on abundance but on other stellar population parameters.
S\'anchez-Bl\'azquez et al. (2003) found that the CN is stronger in Virgo than
in Coma galaxies. The near-IR Ca{\sc II} triplet, which is sensitive to both
the Ca abundance and the IMF (Vazdekis et al. 2003), decreases as a function of
$\sigma$ (Saglia et al. 2002; Cenarro et al. 2003).

\acknowledgments

We thank N. Caon, H.--W. Rix and the referee for useful suggestions, and J.
Blakeslee for providing us with photometric data for NGC~4489 and NGC~7454.
Based on observations made with the European Southern Observatory telescopes
obtained from the ESO/ST-ECF Science Archive Facility.

\end{document}